\documentclass[aps,showpacs,twocolumn]{revtex4}
\usepackage{graphicx}

\begin{document}

\title{Thin accretion disk signatures of slowly rotating
black holes in Ho\v{r}ava gravity}
\author{Tiberiu Harko}
\email{harko@hkucc.hku.hk}
\affiliation{Department of Physics and Center for Theoretical
and Computational Physics,
The University of Hong Kong, Pok Fu Lam Road, Hong Kong}
\author{Zolt\'{a}n Kov\'{a}cs}
\email{zkovacs@hku.hk}
\affiliation{Department of Physics and Center for Theoretical
and Computational Physics,
The University of Hong Kong, Pok Fu Lam Road, Hong Kong}
\author{Francisco S.~N.~Lobo}
\email{flobo@cii.fc.ul.pt}\affiliation{Centro de Astronomia
e Astrof\'{\i}sica da Universidade de Lisboa, Campo Grande, Ed. C8
1749-016 Lisboa, Portugal}

\date{\today}

\begin{abstract}

In the present work, we consider the possibility of observationally testing Ho\v{r}ava gravity by using the accretion disk properties around slowly rotating black holes of the Kehagias-Sfetsos solution in asymptotically flat spacetimes. The energy flux, temperature distribution, the emission spectrum as well as the energy conversion efficiency are obtained, and compared to the standard slowly rotating general relativistic Kerr solution.
Comparing the mass accretion in a slowly rotating Kehagias-Sfetsos geometry in Ho\v{r}ava gravity with the one of a slowly rotating Kerr black hole, we verify that the intensity of the flux emerging from the disk surface is greater for the slowly rotating Kehagias-Sfetsos solution than for rotating black holes with the same geometrical mass and accretion rate. We also present the conversion efficiency of the accreting mass into radiation, and show that the rotating Kehagias-Sfetsos solution provides a much more efficient engine for the transformation of the accreting mass into radiation than the Kerr black holes. Thus, distinct signatures appear in the electromagnetic spectrum, leading to the possibility of directly testing Ho\v{r}ava gravity models by using astrophysical observations of the emission spectra from accretion disks.

\end{abstract}

\maketitle


\section{Introduction}

Recently, Ho\v{r}ava proposed a renormalizable gravity theory in
four dimensions which reduces to Einstein gravity with a
non-vanishing cosmological constant in IR but with improved UV
behaviors \cite{Horava:2008ih,Horava:2009uw}. The latter theory
admits a Lifshitz scale-invariance in time and space, exhibiting a
broken Lorentz symmetry at short scales, while at large distances
higher derivative terms do not contribute, and the theory reduces
to standard general relativity (GR). Since then various properties
and characteristics of the Ho\v{r}ava gravities have been
extensively analyzed, ranging from formal developments
\cite{formal}, cosmology \cite{cosmology}, dark energy
\cite{darkenergy} and dark matter \cite{darkmatter}, and
spherically symmetric solutions
\cite{BHsolutions,Park:2009zra,Lu:2009em,Kehagias:2009is}.
Although a generic vacuum of the theory is anti-de Sitter one,
particular limits of the theory allow for the Minkowski vacuum. In
this limit the leading order the Parameterized post-Newtonian (PPN) coefficients coincide with those of the
pure GR. Thus, the deviations from the conventional GR can be tested
only beyond the post-Newtonian corrections, that is for a system
with strong gravity at astrophysical scales.

There are basically four versions of Ho\v{r}ava gravity, namely, those
with or without the ``detailed balance condition'', and with or without the ``projectability condition''. The ``detailed balance condition'' restricts the form of the potential in the 4--dim Lorentzian action to a specific form in terms of a 3--dim Euclidean theory. In a cosmological context, this condition leads to obstacles, and thus must be abandoned. In this context, the `soft' violation of the `detailed balance' condition modifies the IR behavior. This IR modification, with an arbitrary cosmological constant, represent the analogs of the standard Schwarzschild-(A)dS solutions, which were absent in the original Ho\v{r}ava model. The ``projectability condition'' essentially stems from the fundamental symmetry of the theory, i.e., the foliation-preserving diffeomorphism invariance, and must be respected. Foliation-preserving diffeomorphism consists of a 3--dim spatial and the space-independent time reparametrization.

IR-modified Ho\v{r}ava gravity seems to be consistent with the current observational data \cite{Konoplya:2009ig,Chen:2009eu,Chen:2009bu}, but in order to test its viability more observational constraints are necessary.
In \cite{HaKoLo09b} the possibility of observationally testing Ho\v{r}ava gravity was considered by using the accretion disk properties around black holes. The energy flux, temperature distribution, the emission spectrum as well as the energy conversion efficiency are obtained, and compared to the standard general relativistic case. It was shown that particular signatures can appear in the electromagnetic spectrum, thus leading to the possibility of directly testing Ho\v{r}ava gravity models by using astrophysical observations of the emission spectra from accretion disks. In this paper, we further explore the possibility of testing the viability of Ho\v{r}ava-Lifshitz gravity using thin accretion disk properties, and we generalize the previous results from \cite{HaKoLo09b} to the case of the slowly rotating Ho\v{r}ava-Lifshitz black holes. As for the slowly rotating Ho\v{r}ava-Lifshitz  black hole, we adopt the solution obtained in \cite{Lee:2010xe}, and we carry out an analysis of the properties of the radiation emerging from the surface of the disk. For more information on thin accretion disks around compact objects we refer the reader to Refs. \cite{UrPa95,Miyo95,NoTh73,ShSu73,PaTh74,Th74,Harko:2008vy,HaKoLo09,
Bom,To02,YuNaRe04,Guzman:2005bs,Pun:2008ua,KoChHa09, grav,Harko:2009kj,Pun:2008ae}.

It seems to be an extremely difficult endeavor to find a fully rotating black hole in Ho\v{r}ava-Lifshitz theory as the full equations to be solved are very complicated, so only the slowly rotating regime is analyzed in this work. Note that the ``slowly rotating'' black hole means that one considers up to linear order of the rotating parameter $a=J/M\;(a\ll1)$ in the metric functions, equations of motion, and thermodynamic quantities \cite{Lee:2010xe}. It is also important to emphasize that the slowly rotating Kerr black hole is recovered from the slowly rotating black hole solutions in Ho\v{r}ava gravity, in the IR limit of $\omega \to \infty$.
As compared to the standard general relativistic case, significant differences appear in the energy flux and electromagnetic spectrum for
Ho\v{r}ava black holes, thus leading to the possibility of
directly testing the Ho\v{r}ava-Lifshitz theory by using
astrophysical observations of the emission spectra from accretion
disks.

The present paper is organized as follows. In Sec. \ref{sec:II},
we present the action and the specific solution of slowly rotating black holes. In Sec. \ref{sec:III}, we present some essential thermal equilibrium radiation properties of thin accretion disks in stationary axisymmetric spacetimes. In Sec. \ref{sec:IV}, we analyze the basic properties of matter forming a thin accretion disk around vacuum black holes in Ho\v{r}ava gravity, and compare the results with the slowly rotating Kerr solution. We discuss and conclude our results in Sec. \ref{sec:concl}. Furthermore, for self-completeness and self-consistency, we review the formalism and the physical properties of the thin disk accretion onto compact objects in Appendix \ref{App1}.

\section{Slowly rotating black holes in Ho\v{r}ava gravity}
\label{sec:II}

In this section, we briefly review the Ho\v{r}ava-Lifshitz theory,
where differential geometry of foliations represents the proper
mathematical setting for the class of gravity theories studied by
Ho\v{r}ava \cite{Horava:2009uw}. Thus, it is useful to use the ADM formalism, where the four-dimensional metric is parameterized by the following
\begin{equation}
ds^2=-N^2c^2\,dt^2+g_{ij}\left(dx^i+N^i\,dt\right)
\left(dx^j+N^j\,dt\right)\,,
\end{equation}
where $N$ is the lapse function, $N^i$, the shift vector, and $g_{ij}$ the 3-dimensional spatial metric.

In this context, the Einstein-Hilbert action is given by
\begin{equation}
S=\frac{1}{16\pi G}\int d^4x \;\sqrt{g}\,N\left(K_{ij}K^{ij}-
K^2+R^{(3)}-2\Lambda \right) , \label{EHaction}
\end{equation}
where $G$ is Newton's constant, $R^{(3)}$ is the three-dimensional
curvature scalar for $g_{ij}$. The extrinsic curvature, $K_{ij}$,
is defined as
\begin{equation}
K_{ij}=\frac{1}{2N}\left(\dot{g}_{ij}-\nabla_iN_j-\nabla_jN_i\right),
\end{equation}
where the dot denotes a derivative with respect to $t$, and
$\nabla_i$ is the covariant derivative with respect to the spatial
metric $g_{ij}$.

The IR-modified Ho\v{r}ava action is given by
\begin{eqnarray}
S&=&\int dt\,d^3x
\;\sqrt{g}\,N\Bigg[\frac{2}{\kappa^2}\left(K_{ij}K^{ij}-\lambda
K^2\right) -\frac{\kappa^2}{2\nu^4} C_{ij}C^{ij}
     \nonumber  \\
&&+\frac{\kappa^2\mu}{2\nu^2}\epsilon^{ijk}R^{(3)}_{il}\nabla_j
R^{(3)l}{}_{k}-\frac{\kappa^2\mu^2}{8}R^{(3)}_{ij}R^{(3)ij}
    \nonumber  \\
&&    +\frac{\kappa^2\mu^2} {8(3\lambda-1)}
\left(\frac{4\lambda-1}{4}(R^{(3)})^2-\Lambda_W
R^{(3)}+3\Lambda_W^2\right)
    \nonumber   \\
&& +\frac{\kappa^2\mu^2\omega}{8(3\lambda-1)} R^{(3)}\Bigg],
\label{Haction}
\end{eqnarray}
where $\kappa$, $\lambda$, $\nu$, $\mu$, $\omega$ and $\Lambda_W$
are constant parameters. $C^{ij}$ is the Cotton tensor, defined as
\begin{equation}
C^{ij}=\epsilon^{ikl}\nabla_k\left(R^{(3)j}{}_{l}-\frac{1}{4}
R^{(3)}\delta^j_{l}\right).
\end{equation}
The last term in Eq.~(\ref{Haction}) represents a `soft'
violation of the `detailed balance' condition, which modifies the
IR behavior. This IR modification term, $\mu^2 R^{(3)}$,
generalizes the original Ho\v{r}ava model (we have used the
notation of Ref. \cite{Park:2009zra}).

The fundamental constants of the speed of light $c$, Newton's
constant $G$, and the cosmological constant $\Lambda$ are defined
as
\begin{equation}
c^2=\frac{\kappa^2\mu^2|\lambda_W|}{8(3\lambda-1)^2}\quad
G=\frac{\kappa^2c^2}{16\pi(3\lambda-1)}\quad
\Lambda=\frac{3}{2}\Lambda_W c^2.
\end{equation}

Consider the static and spherically symmetric metric given by
\begin{equation}
ds^2=-N^2(r)\,dt^2+\frac{dr^2}{f(r)}+r^2 \,(d\theta ^2+\sin
^2{\theta} \, d\phi ^2) \label{SSSmetric},
\end{equation}
where $N(r)$ and $f(r)$ are arbitrary functions of the radial
coordinate, $r$.

Imposing the specific case of $\lambda=1$, $\beta=4\omega M$ ($M$ is the mass parameter) and $\Lambda_W=0$, one obtains the Kehagias and Sfetsos's (KS)
asymptotically flat solution \cite{Kehagias:2009is}, given by
\begin{equation}
f_{\rm KS} =N^2_{\rm KS}= 1 + \omega r^2 \left (1-\sqrt{1 +
\frac{4 M}{\omega r^3}}\right )\,.
  \label{KSsolution}
\end{equation}
In the limit of $\omega \to \infty$, it reduces to the Schwarzschild form.

Note that there is an outer (event) horizon, and an inner (Cauchy) horizon at
\begin{equation}
r_{\pm}=M\left[1\pm\sqrt{1-1/(2\omega M^2)}\right].
\end{equation}
To avoid a naked singularity at the origin, one also needs to
impose the condition
\begin{equation}
\omega M^2\geq \frac{1}{2}.
\end{equation}
Note that in the GR regime, i.e., $\omega M^2 \gg 1$, the outer
horizon approaches the Schwarzschild horizon, $r_+\simeq 2M$, and
the inner horizon approaches the central singularity, $r_- \simeq
0$.

In the present work we propose to study the thin accretion
disk models applied for slowly rotating black holes in Ho\v{r}ava-Lifshitz gravity models \cite{Lee:2010xe}, and carry out an analysis of the properties of the radiation emerging from the surface of the disk. It does indeed seem to be a formidable task to find a fully rotating black hole in Ho\v{r}ava-Lifshitz theory as the full equations to be solved are very complicated, so only the slowly rotating regime is analyzed in this work. Note that the ``slowly rotating'' black hole means that one considers up to linear order of the rotating parameter $a=J/M\;(a\ll1)$ in the metric functions, equations of motion, and thermodynamic quantities. In this work we use the results obtained in Ref. \cite{Lee:2010xe}, where the slowly rotating black hole is interpreted as arising from the breaking of spherical to axial symmetry. Rather than reproduce the analysis carried out, we refer to reader to Ref. \cite{Lee:2010xe} for details.

Thus, the slowly rotating black hole solution obtained in \cite{Lee:2010xe} is given by
\begin{eqnarray}
ds^2_{\rm slow~KS} &=& -f_{\rm KS}(r) dt^2 + \frac{dr^2}{f_{\rm
KS}(r)} + r^2 d \theta^2
   \nonumber  \\
&&+r^2 \sin^2 \theta \left ( d \phi^2 - \frac{4J}{r^3} dt d\phi \right )\,,
\label{slowrsol}
\end{eqnarray}
where the factor $f_{\rm KS}(r)$ reduces to the KS solution, and is given by
Eq. (\ref{KSsolution}).

For self-completeness and self-consistency, the Kerr black hole is also presented
\begin{equation}
ds^2_{\rm Kerr} = - \frac{\rho^2 \Delta_r}{\Sigma^2} dt^2 +
\frac{\rho^2}{\Delta_r} dr^2
     + \rho^2 d\theta^2 + \frac{\Sigma^2 \sin^2 \theta}{\rho^2}
       \left ( d\phi - \xi dt \right )^2,
\label{kerr_metric1}
\end{equation}
where
\begin{eqnarray}
\rho^2 &=& r^2 + a^2 \cos^2 \theta, \nonumber \\
\Delta_r &=& \left ( r^2 + a^2 \right ) - 2 M r, \nonumber \\
\Sigma^2 &=& \left ( r^2 + a^2 \right )^2 - a^2 \sin^2 \theta \Delta_r, \nonumber \\
\xi &=& \frac{2Mar}{\Sigma^2}.
\end{eqnarray}
In the slowly rotating limit of $J \ll M (a \ll 1)$, the Kerr
solution reduces to
\begin{eqnarray} \label{SK} ds^2_{\rm
slow~Kerr}&=&-\Big(1-\frac{2M}{r}\Big)dt^2+\frac{dr^2}{\Big(1-\frac{2M}{r}\Big)}
+r^2d\theta^2
   \nonumber \\
&&+r^2\sin^2\theta\Big(d\phi^2-\frac{4J}{r^3}dt
d\phi\Big)\label{kerr_metric}
\end{eqnarray}

It was argued in \cite{Lee:2010xe} that the slowly rotating Kerr black hole could be interpreted as arising from the breaking of spherical to axial symmetry.  It is easily checked that in the limit of $\omega \to \infty$, the slowly rotating solution (\ref{slowrsol}) leads to the slowly rotating Kerr solution (\ref{SK}). In the analysis of the thin disk properties that follow, we compare the geometries given by the metrics (\ref{slowrsol}) and (\ref{kerr_metric}), respectively.

\section{Geodesic motion of test particles in slowly rotating Kehagias-Sfetsos geometry}\label{sec:III}


In this section we consider the thermal radiation properties of thin accretion disks in stationary axisymmetric spacetimes. The formalism has been extensively presented in the literature. However, in order to analyze the electromagnetic signatures of thin accretion disks around slowly rotating black holes, for self-completeness and self-consistency, we consider the main results of stationary and axially symmetric spacetimes in this section and the physical properties of accretion disks in Appendix \ref{App1}.

The physical properties and the electromagnetic radiation
characteristics of particles moving in circular orbits around
compact bodies are determined by the geometry of the
spacetime around the compact object.  For a stationary and axially
symmetric geometry the metric is given in a general form by
\begin{equation}\label{rotmetr1}
ds^2=g_{tt}\,dt^2+2g_{t\phi}\,dt d\phi+g_{rr}\,dr^2
+g_{\theta\theta}\,d\theta^2+g_{\phi\phi}\,d\phi^2\,.
\end{equation}
In the equatorial approximation, i.e., $|\theta-\pi /2|\ll 1$, which is the case of interest for our analysis, the metric functions $g_{tt}$, $g_{t\phi}$, $g_{rr}$, $g_{\theta\theta}$ and $g_{\phi\phi}$ only depend on the radial coordinate $r$.

To compute the relevant physical quantities of the thin accretion
disks, we determine first the radial dependence of the angular
velocity $\Omega $, of the specific energy $\widetilde{E}$, and of
the specific angular momentum $\widetilde{L}$, respectively,  of particles moving
in circular orbits in a stationary and axially symmetric geometry. The geodesic equations of the motion take the following form
\begin{eqnarray}
\frac{dt}{d\tau}&=&\frac{\widetilde{E}
g_{\phi\phi}+\widetilde{L}g_{t\phi}}{g_{t\phi}^2-g_{tt}g_{\phi\phi}},
   \label{geodeqs1}   \\
\frac{d\phi}{d\tau}&=&-\frac{\widetilde{E}
g_{t\phi}+\widetilde{L}g_{tt}}{g_{t\phi}^2-g_{tt}g_{\phi\phi}},
    \label{geodeqs2}  \\
g_{rr}\left(\frac{dr}{d\tau}\right)^2&=&-1+\frac{\widetilde{E}^2
g_{\phi\phi}+2\widetilde{E}\widetilde{L}g_{t\phi}
+\widetilde{L}^2g_{tt}}{g_{t\phi}^2-g_{tt}g_{\phi\phi}}.
    \label{geodeqs3}
\end{eqnarray}

From Eq.~(\ref{geodeqs3}) one can introduce an effective potential term as
\begin{equation}\label{roteffpot}
V_{eff}(r)=-1+\frac{\widetilde{E}^2
g_{\phi\phi}+2\widetilde{E}\widetilde{L}g_{t\phi}
+\widetilde{L}^2g_{tt}}{g_{t\phi}^2-g_{tt}g_{\phi\phi}}.
\end{equation}

\begin{figure*}[t]
\centering
\includegraphics[width=.48\textwidth]{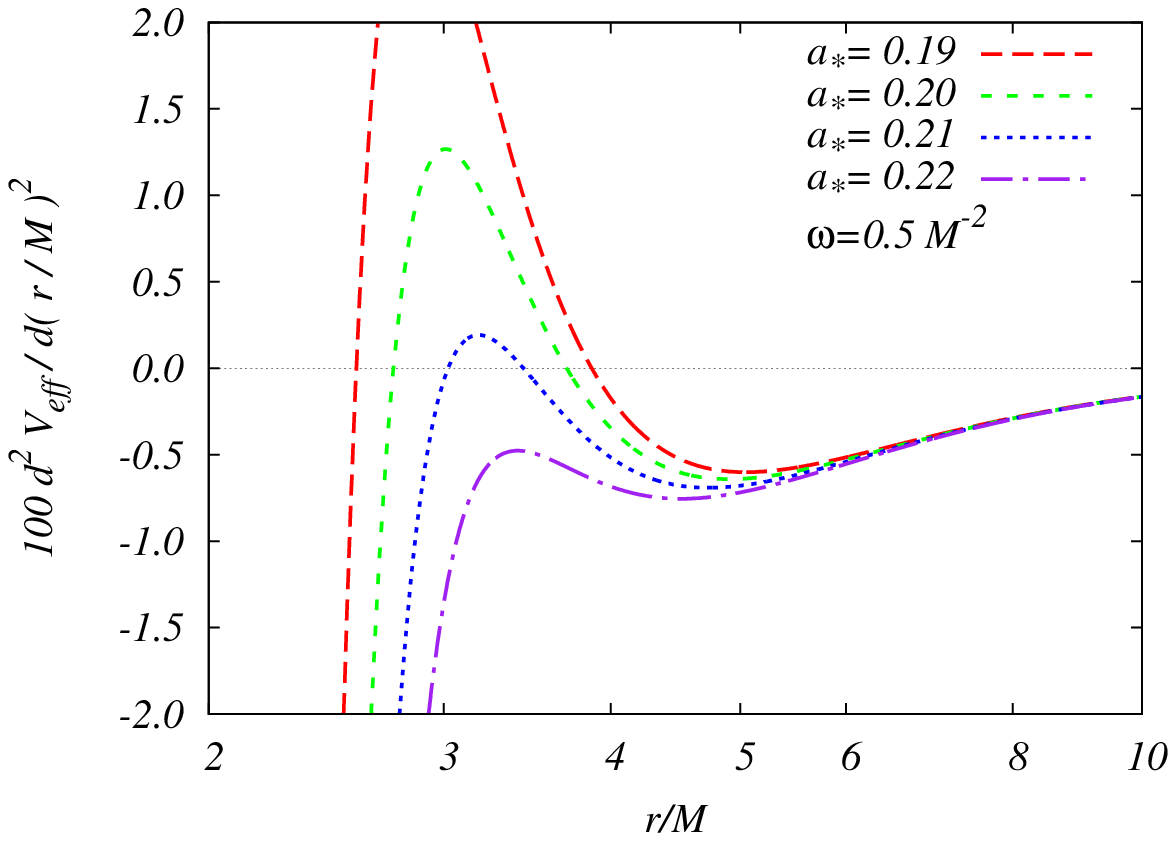}
\includegraphics[width=.48\textwidth]{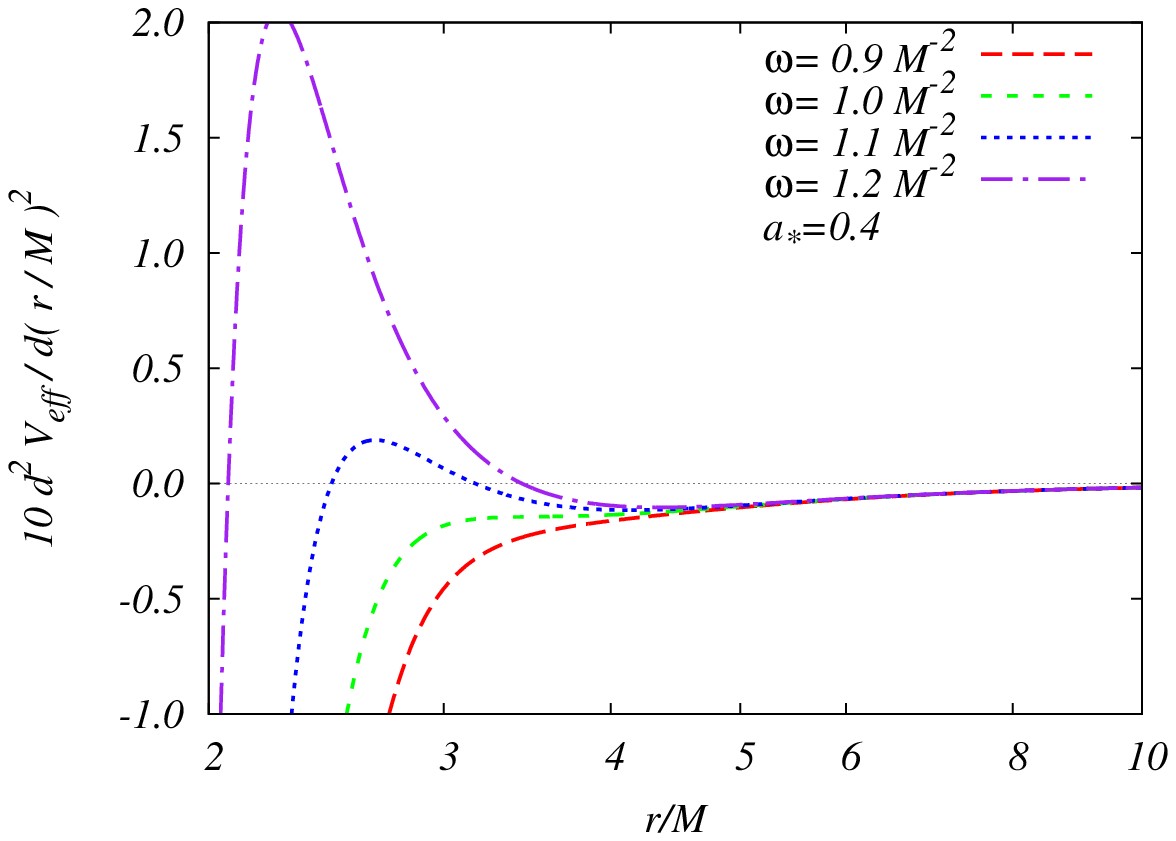}
\caption{The second order derivative of the effective potential with respect to $(r/M)$ for a slowly rotating KS black hole with total mass $M$ for different values of the parameters $\omega$ and $a_*$.}
\label{fig0}
\end{figure*}

For stable circular orbits in the equatorial plane the following
conditions must hold: $V_{eff}(r)=0$ and $V_{eff,\;r}(r)=0$, where
the comma in the subscript denotes a derivative with respect to
the radial coordinate $r$. These conditions provide the specific
energy, the specific angular momentum and the angular velocity of
particles moving in circular orbits for the case of spinning
general relativistic compact spheres, given by
\begin{eqnarray}
\widetilde{E}&=&-\frac{g_{tt}+g_{t\phi}\Omega}{\sqrt{-g_{tt}
-2g_{t\phi}\Omega-g_{\phi\phi}\Omega^2}},
    \label{rotE}  \\
\widetilde{L}&=&\frac{g_{t\phi}+g_{\phi\phi}\Omega}{\sqrt{-g_{tt}
-2g_{t\phi}\Omega-g_{\phi\phi}\Omega^2}},
     \label{rotL}  \\
\Omega&=&\frac{d\phi}{dt}=\frac{-g_{t\phi,r}+\sqrt{(g_{t\phi,r})^2
-g_{tt,r}g_{\phi\phi,r}}}{g_{\phi\phi,r}}.
     \label{rotOmega}
\end{eqnarray}

The marginally stable orbit around the central object can be
determined from the condition $V_{eff,\;rr}(r)=0$. To this effect,
we formally represent the effective potential as
\[
V_{eff}(r)\equiv-1+\frac{f}{g},
\]
where
\begin{eqnarray*}
f & \equiv & \widetilde{E}^{2}g_{\phi\phi}
+2\widetilde{E}\widetilde{L}g_{t\phi}+\widetilde{L}^{2}g_{\phi\phi},\\
g & \equiv & g_{t\phi}^{2}-g_{tt}g_{\phi\phi},
\end{eqnarray*}
and where the condition $g\neq 0$ is imposed. From $V_{eff}(r)=0$,
we obtain first $f=g$. The condition $V_{eff,\;r}(r)=0$ provides
$f_{,r}g-fg_{,r}=0$. Thus, from these conditions one readily
derives $V_{eff,\;rr}(r)=0$, which provides the following
important relationship
\begin{eqnarray}
 0& = & (g_{t\phi}^2-g_{tt}g_{\phi\phi})V_{eff,rr}\nonumber\\
 & = &  \widetilde{E}^{2}g_{\phi\phi,rr}+2\widetilde{E}\widetilde{L}g_{t\phi,rr} +\widetilde{L}^{2}g_{tt,rr}  \nonumber \\
& & -(g_{t\phi}^{2} -g_{tt}g_{\phi\phi})_{,rr}\;,\label{mso-r}
\end{eqnarray}
where $g_{t\phi}^2-g_{tt}g_{\phi\phi}$ (appearing as a cofactor in
the metric determinant) never vanishes. By inserting
Eqs.~(\ref{rotE})-(\ref{rotOmega}) into Eq.~(\ref{mso-r}) and
solving this equation for $r$, we obtain the radii of the
marginally stable orbits, once the metric coefficients $g_{tt}$,
$g_{t\phi}$ and $g_{\phi\phi}$ are explicitly given.

In the context of the slowly rotating KS black hole given by the metric (\ref{slowrsol}), and considering the equatorial approximation ($|\theta-\pi/2|\ll1$) of the geometry, the particles moving in Keplerian orbits around the slowly rotating black hole have the following specific energy, specific angular momentum and angular velocity
\begin{eqnarray}
\widetilde{E}&=&\frac{1+\omega r^2  + 2M^2a_* r^{-1} - \omega  h r^2 }{\sqrt{1+(\omega-\Omega^2)r^2+ 4 Ma_*\Omega r ^{-1}-\omega  h r^2 }}\:,\label{rotE2}\\
\widetilde{L}&=&\frac{r^2\Omega  - 2M^2a_*r^{-1}}{\sqrt{1+(\omega-\Omega^2)r^2 + 4 Ma_*\Omega r ^{-1}-\omega  hr^2}}\:,\label{rotL2}\\
\Omega&=&\frac{\sqrt{h(M^4a^2_*+\omega r^6)-Mr^3-\omega r^6}-2M^2a_*\sqrt{h}}{2r^3\sqrt{h}}\label{rotOmega2}
\end{eqnarray}
respectively, where $a_*=J/M^2$ and $h=\sqrt{1+4M/(\omega r^3)}$ are defined for notational simplicity. By inserting these equations into the second order derivative (\ref{mso-r}) of the effective potential, we can determine the radii of the marginally stable orbits for different values of the spin parameter $a_*$ and of the parameter $\omega$. In order to obtain consistency with the metric, Eqs.~(\ref{rotE2})-(\ref{rotOmega2}) should be linearized with respect to the parameter $a_*=J/M^2$. However, since $\widetilde{E}$, $\widetilde{L}$ and $\Omega $ are constants of motion, their numerical values are also obtained from the particular values given to $a_*$ and $\omega $, respectively. Therefore any series expansion or linearization of Eqs.~(\ref{rotE2})-(\ref{rotOmega2}) will just slightly modify the numerical values of the constants of motion, without significantly affecting the physical results.

In Fig.~\ref{fig0} we plot the quantity $d^2 V_{eff} / d (r/M)^2$ as a function of the dimensionless radius $r/M$.
The left plot shows the radial profile of the second order derivative calculated for different values of $a_*$ in a narrow range between 0.19 and 0.22 at a fixed value of $\omega$ (set to its the minimal value $1/2M^2$). For $a_*=0.19$, 0.20 and 0.21, $d^2 V_{eff} / d (r/M)^2$ has zeroes, and the larger value of the root provides the radius $r_{ms}$ of the marginally stable orbit. But the second derivative of the effective potential, calculated for $a_*=0.22$, remains negative for any $r>r_{+}$. Hence stable circular orbits do exist for $a_*=0.22$ and $\omega=1/2M^2$ in the entire equatorial plane outside the horizon ($r_{\pm}=M$). The critical value of $a_*$ above which no marginally stable orbit exists depends on $\omega$. The right hand side plot in Fig.~\ref{fig0} shows that a given value of $a_*$ becomes critical if we decrease $\omega$: for $a_*=0.4$, the second order derivative of $V_{eff}$, calculated at $\omega=1.1 M^{-2}$ and $1.2 M^{-2}$, still has some roots, but there
  are no marginally stable orbits for $\omega=1 M^{-2}$ and $\omega=0.9 M^{-2}$, respectively. The presence of the
critical value for $a_*$, and its numerical value, certainly depends on the approximation used to obtain the metric. Such a critical value may indicate a breakdown of the metric for enough high rotational velocities. For an exact solution and for high spins, the location of the stable circular orbits may be different. However, since the solution we are considering is valid for small $a_*$ only, the extrapolation of our results to higher spin values may lead to unphysical results.

\begin{figure}
\centering
\includegraphics[width=.48\textwidth]{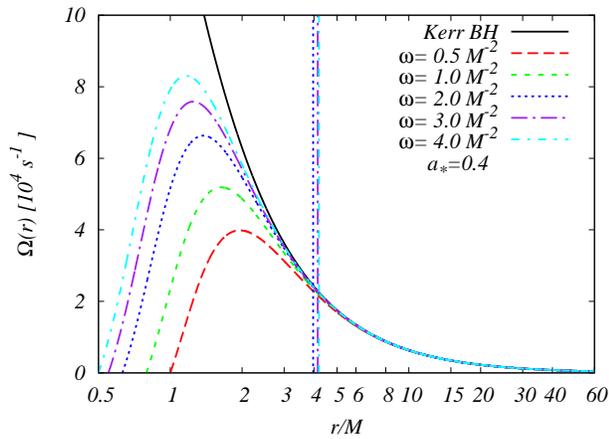}
\caption{The rotational velocity of the Keplerian orbits in the equatorial plane as a function of $(r/M)$ for a slowly rotating KS black hole with total mass  $1M_{\odot}$ for different values of the parameters $\omega$ and $a_*=0.4$. The vertical lines indicate the location of the ISCO for those discs for which it exists.}
\label{fig00}
\end{figure}

Another interesting property of the Keplerian motion around a slowly rotating KS black hole is the radial dependence of the orbital frequency of the particles, shown in Fig.~\ref{fig00}. In Fig.~\ref{fig00} we plot $\Omega$ versus $r/M$ for a fixed value of $a_*=0.4$,  and for different values of $\omega$. With decreasing values of $\omega$, the radial profiles of the angular velocity decrease at lower radii, as compared to the general relativistic case. As we approach the horizon, this decrease is not only a relative decrease, but the curves fall to zero, instead of following the increase of the general relativistic case. Hence $\Omega$ has a maximal value for each pair of the parameters $a_*$ and $\omega$, as opposed to the angular velocity of the particles rotating around a Kerr black hole, where $\Omega$ remains a monotonous increasing function as we approach the horizon.

\section{Electromagnetic signatures of accretion
disks around slowly rotating KS black holes}\label{sec:IV}

\begin{figure*}[t]
\centering
\includegraphics[width=.48\textwidth]{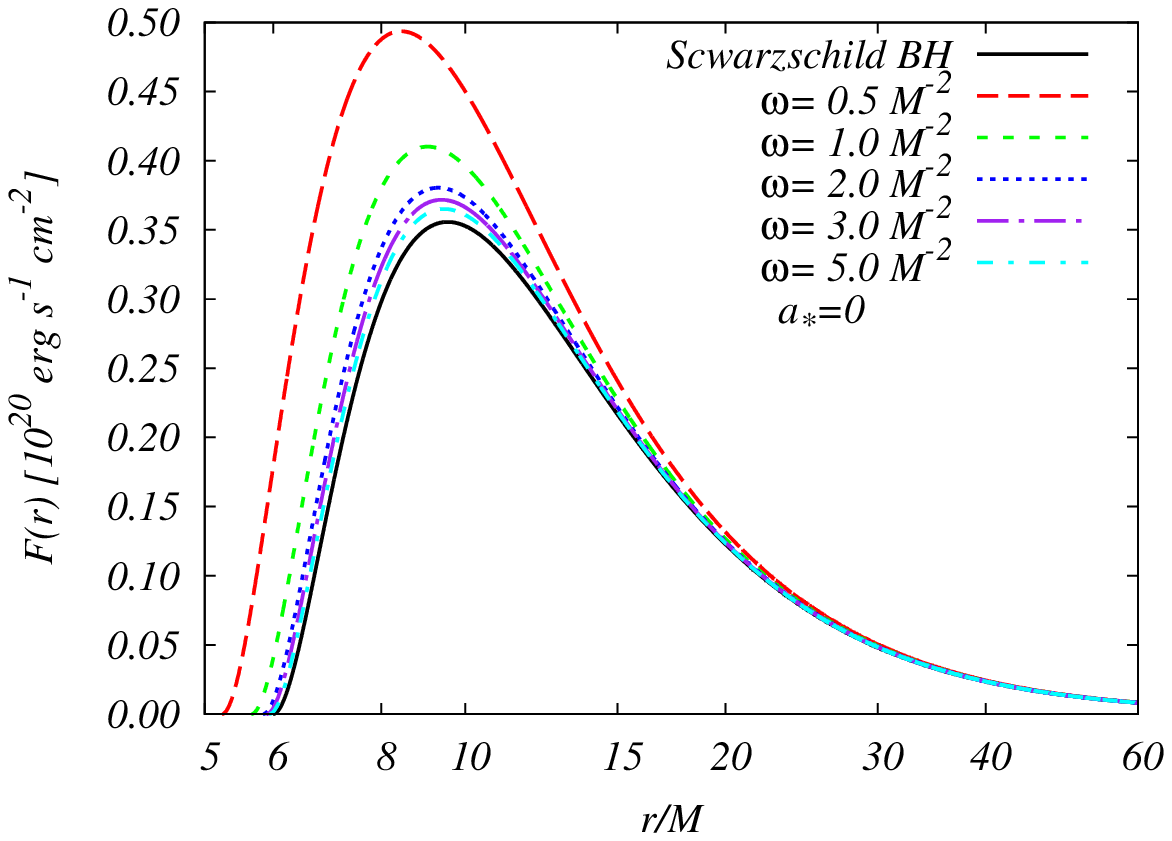}
\includegraphics[width=.48\textwidth]{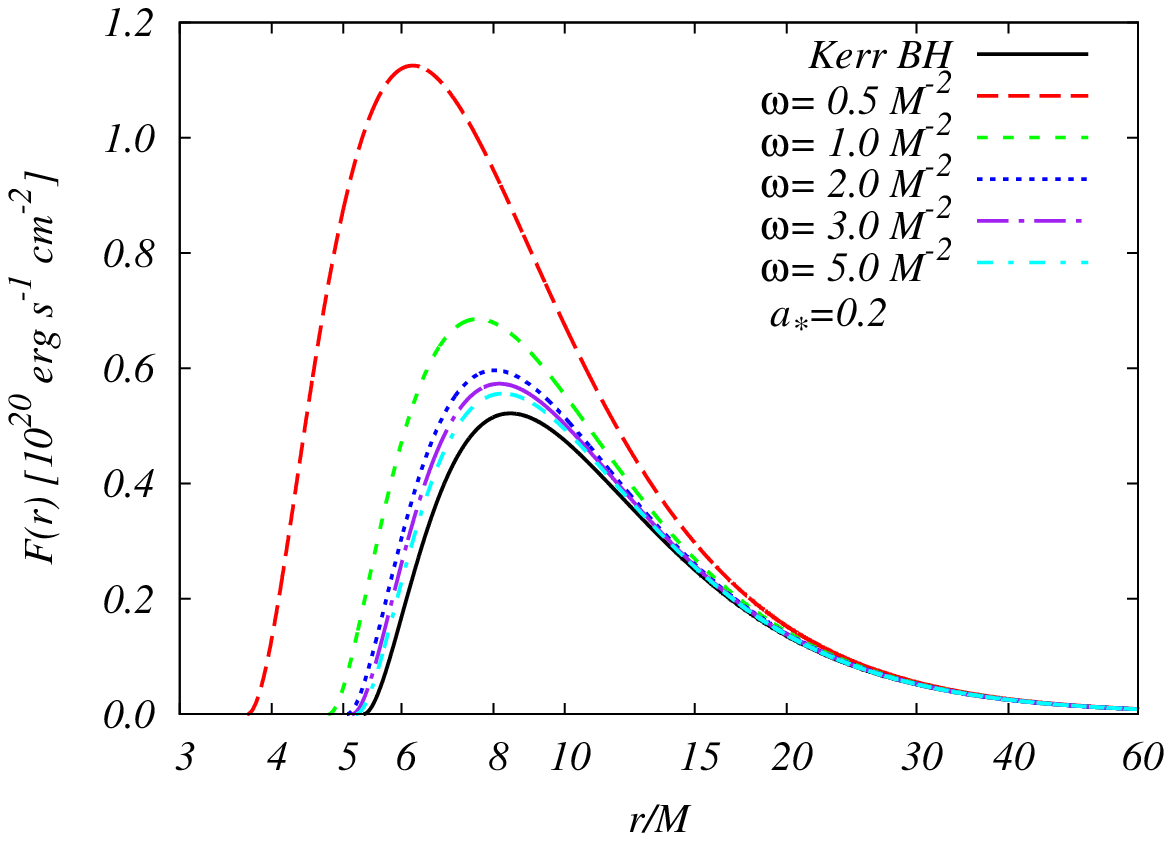}\\
\includegraphics[width=.48\textwidth]{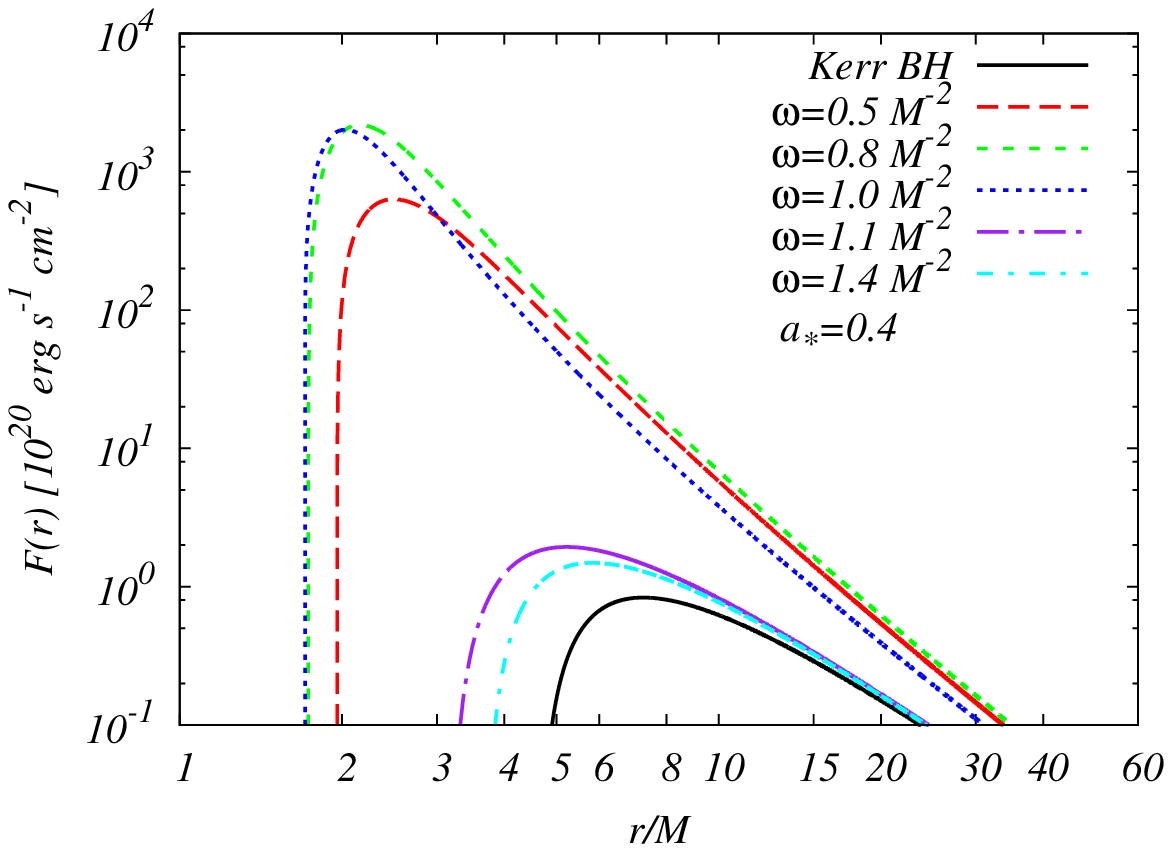}
\includegraphics[width=.48\textwidth]{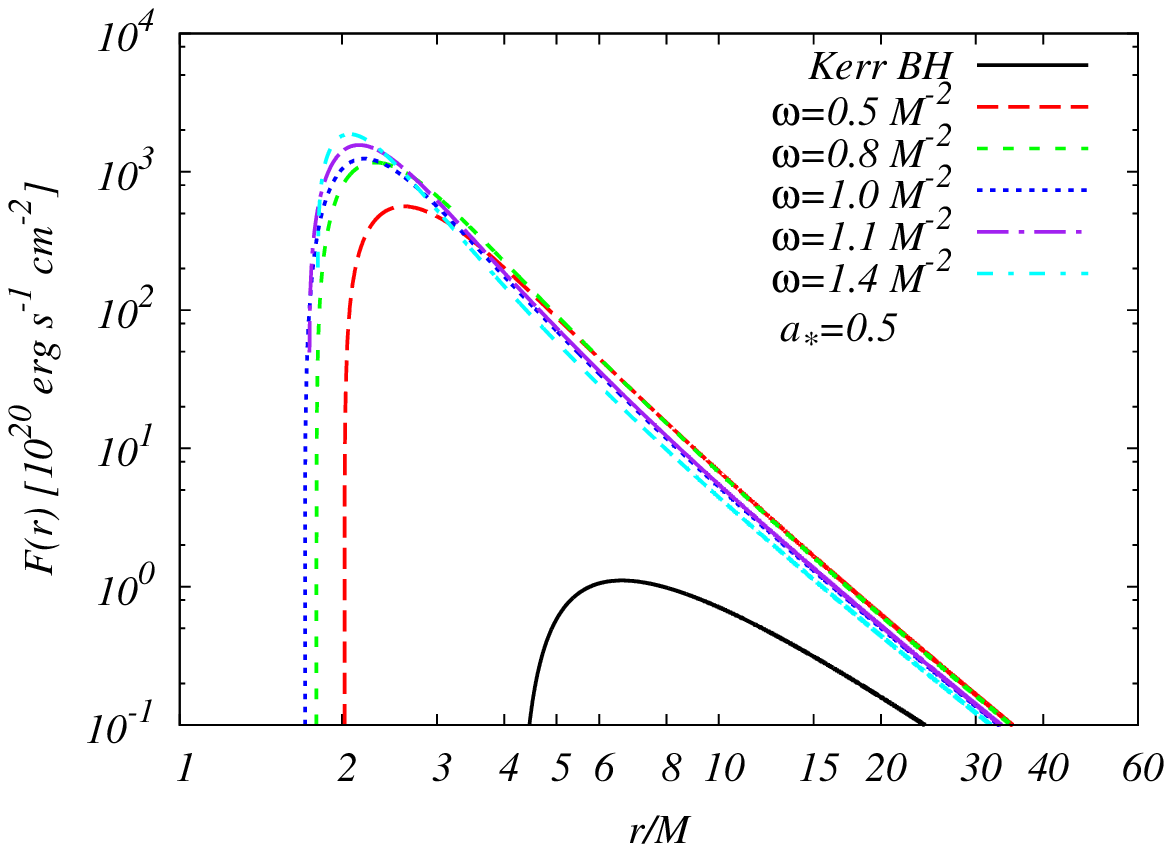}
\caption{The energy flux $F(r)$ radiated by the disk around a static and a slowly rotating KS black hole with the total mass of $1M_{\odot}$ for different values of the parameters $\omega$ and $a_*$. The mass accretion rate is set to $10^{-12}M_{\odot}$/yr.}
\label{fig1}
\end{figure*}

In this section we analyze the electromagnetic signatures of accretion disks around the slowly rotating KS black hole. We rely heavily on the formalism of the physical properties of thin accretion disks outlined in Appendix \ref{App1}.

In Fig.~\ref{fig1} we present the time-averaged fluxes of radiant energy $F(r)$ released from the accretion disks rotating around slowly rotating KS black holes. The fluxes were calculated by using Eqs. (\ref{rotE2})-(\ref{rotOmega2}) in the flux integral Eq.~(\ref{F}). The fluxes depend on two basic parameters, namely, the dimensionless spin parameter of the black hole, defined by $a_*=J/M^2$, and the free parameter $\omega$, which appears in the solution given by Eq.~(\ref{KSsolution}). For a fixed total mass, $M=1M_{\odot}$, and accretion rate, $\dot{M}_0=10^{-12}M_{\odot}$/yr, in Fig.~\ref{fig1} we present the radial distributions of $F$ for different values of $a_*$ and $\omega$. For comparison, we also plot the flux profiles for accretion disks around general relativistic black holes with the corresponding mass and spin parameter. The first plot shows the static case, which was already discussed in \cite{HaKoLo09b}. With the decreasing value of the parameter $\omega$, the deviation of the spacetime geometry of the KS black hole from the geometry of the Schwarzschild black hole is enhanced. As a result, the flux profiles exhibit higher and higher differences. For low values of $\omega$ we obtain a higher maximal flux, as the inner edge of the accretion disk is shifted to lower radii (left edges of the flux profiles in the plots). The location of the flux maximum has a similar shift: with decreasing $\omega$, $F_{max}$ is shifted to lower radii. By comparing the flux maxima  we see that $F_{max}$ is about 1.4 times higher for the minimal value of $\omega$ than the maximum derived for the Schwarzschild black hole ($\omega\rightarrow\infty$).

Similar trends can be found for the slowly rotating cases shown in the rest of the plots of Fig.~\ref{fig1}. For $a_*=0.2$, the configurations with decreasing values of $\omega$ produce higher and higher flux values. The increase in the fluxes are again due to the increase in the surface of the accretion disk around the rotating KS black holes. These increase is clearly indicated by the large shifts of the inner disk edge to lower radii.
Comparing these figures with the static case, both the shift of the inner edge and the enhancement in the flux amplitudes are much higher for the same values of $\omega$: for $\omega=0.5M^{-2}$, the maximal flux is already more than two times higher than $F_{max}$ derived for the slowly rotating Kerr solution. By further increasing $a_*$, we can also increase the effect of the variation of $\omega$ on the radial structure of the standard thin accretion disk. We have seen that there is a critical value of the spin parameter for each $\omega$, above which no marginally stable orbits can exist outside the horizon of the Kehagias-Safest black holes. If $a_*$ exceeds this critical value, then the inner edge of the accretion disk is already located at the horizon, and the disk surface is considerably increased, as compared to the configuration with the spin parameter somewhat lower than the critical value. As a result, the amplitudes of the integrated flux have an enormous increase, i.e, there is a strong enhancement in the thermal radiation released from the disk surface.

A deeper physical interpretation of the breakdown of the standard accretion disk model for $<r_{max}$ may be associated with these high flux values.
Some composite accretion disk models consider a geometrically thick and optically thin hot corona,  positioned between the
marginally stable orbit and the inner edge of the geometrically thin, and
optically thick, accretion disk, where an inner edge is lying at a few gravitational radii \cite{ThPr75}.
In other type of composite models, the corona lies above and under the
accretion disk, and the soft photons, arriving from the disk, produce a hard
emission via their inverse comptonization by the thermally hot electrons in
the corona \cite{LiPr77}. In both configurations, the disk is
truncated at several gravitational radii - reducing the soft photon flux of
the disk - and the soft and hard components of the broad band X-ray
spectra of black holes were attributed to the thermal radiation of
the accretion disk and the emission mechanism in the corona, respectively.
For KS black holes the innermost region of the disk with $r<r_{max}$ may indicate the
presence of a hot corona where the rest mass accretion together with the energy and the angular
momentum transport from the disk for $r>r_{max}$ to the region of the corona at lower radii cannot be described with the geometrically thin disk scheme. Thermal cooling of the corona via radiation is not able to maintain a thermodynamical equilibrium and the transport processes are dominated by heat conduction in the outer region of the corona or advection in a geometrically thick disk  between $r_+$ and $r_{max}$ in the core of the corona. In order to give a detailed description of the mechanism driving matter, energy and angular momentum into the black hole in the innermost region, one should construct at least a semi-analytic model without vertical averaging of the physical quantities of the accretion disk.

Here we carry out the analysis of the presence of the critical behavior only in the framework of the standards Novikov-Thorne disk models, which can still provide useful information about the whole phenomenon.
This effect can be seen in the last two plots in Fig.~\ref{fig1}. Since for $\omega>1.1 M^{-2}$ the value $a_*=0.4$ is still under the critical value of the spin parameter, the maximal flux has still a moderate increase. If we set $\omega$ to $1.0 M^{-2}$ or higher values, $a_*=0.4$ exceeds the critical spin value shown in Fig.~\ref{fig0}, and the left edge of the flux profile is located at radii $\le2r/M$. Nevertheless, these radii are still much higher than $r_{+}$ obtained for the corresponding values of $\omega$. This is due to the fact that the radius $r_{max}$ at which $\Omega $ attains its maximal value is greater than $r_+$. For lower radii $\Omega$ is already decreasing, as shown in Fig.~\ref{fig00}, and from Eq.~(\ref{F}) we obtain negative flux values. As a possible physical interpretation we can state that although the inner disk edge can reach the horizon, in the region between $r_{+}$ and $r_{max}$ no physical mechanism is available for the radiative cooling of the disk, and thermal photons cannot leave the disk surface. Close to the inner edge of the disk, other processes might play a role in cooling the disk, and the standard thin disk model likely breaks down in this region. For $r>r_{max}$ the latter model still seems to be a satisfying description, and thus we can assume that the disk is in the state of the thermal equilibrium or at least very close to it.

If for a given $\omega$ the spin exceeds the critical value,  then the minimal radius of the radiating disk surface is determined by the location of the Keplerian orbit with the maximal rotational frequency. As Fig.~\ref{fig00} shows, the value of the radial coordinate at which $\Omega$ is maximal is proportional to $\omega$. Hence, this radius is maximal for the minimal value of $\omega$, and the disk surface emitting thermal photons is minimal. Then, above the critical spin value, the relation between the flux maxima and the value of the parameter $\omega$ is inverted in comparison with the trends found below the critical spin value. This means that the higher the value of $\omega$ (the smaller the deviation from the general relativistic geometry), the higher the value of the maximal flux, and the lower the radial coordinate of the inner edge of the radiating zone.

At the limit $a_*=0.5$ of the slow rotating regime, the flux maximum is three orders of magnitudes higher for $\omega=1.4M^{-2}$ than the maximal value of $F(r)$ obtained for Kerr black holes. For all the values of $\omega$ presented in the last plot the spin value is higher than the critical value and each configuration of the thin disk rotating around the KS black hole emits much more thermal photons than the disk rotating around the Kerr black hole do. This holds even for the minimal value of $\omega$. For $\omega\gg1$ we can still reach the limit above which $a_*$ is no longer greater than the critical spin value, and we can approach the lower flux values, which are typical for general relativistic black holes.

\begin{figure*}[t]
\centering
\includegraphics[width=.48\textwidth]{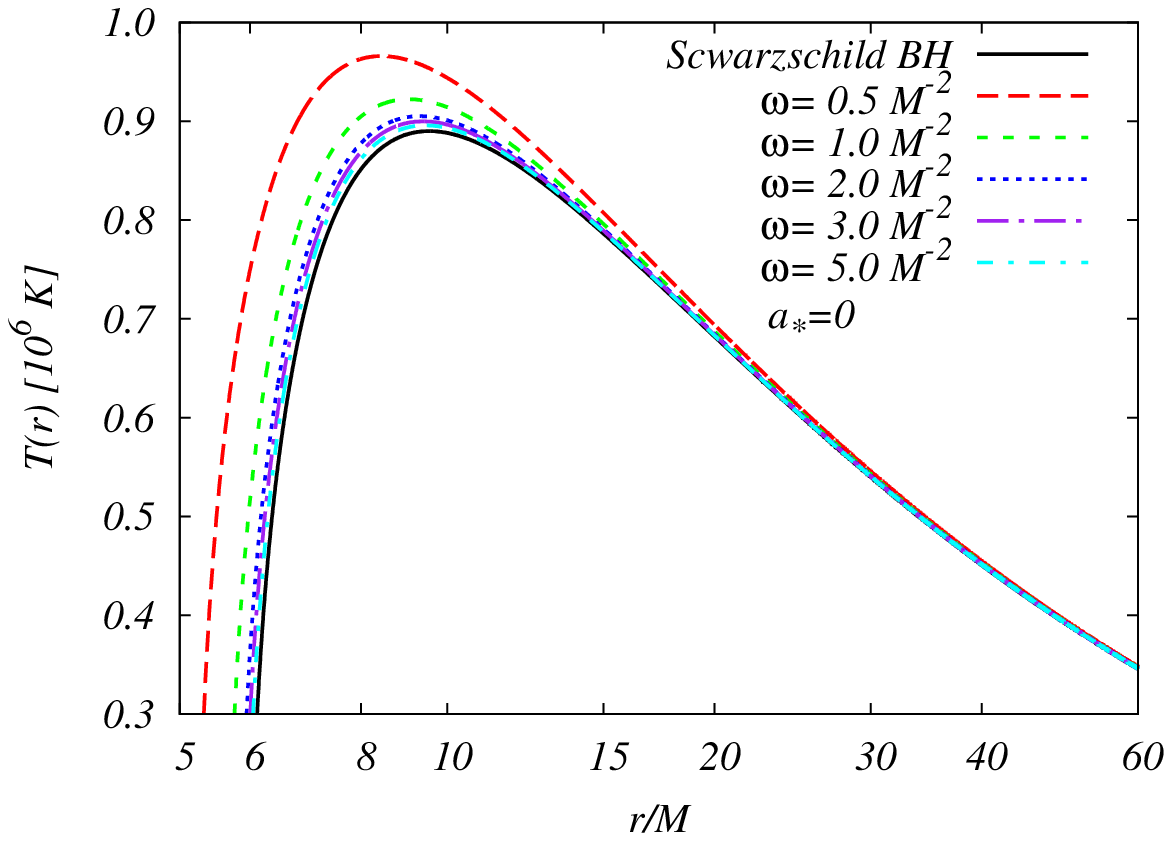}
\includegraphics[width=.48\textwidth]{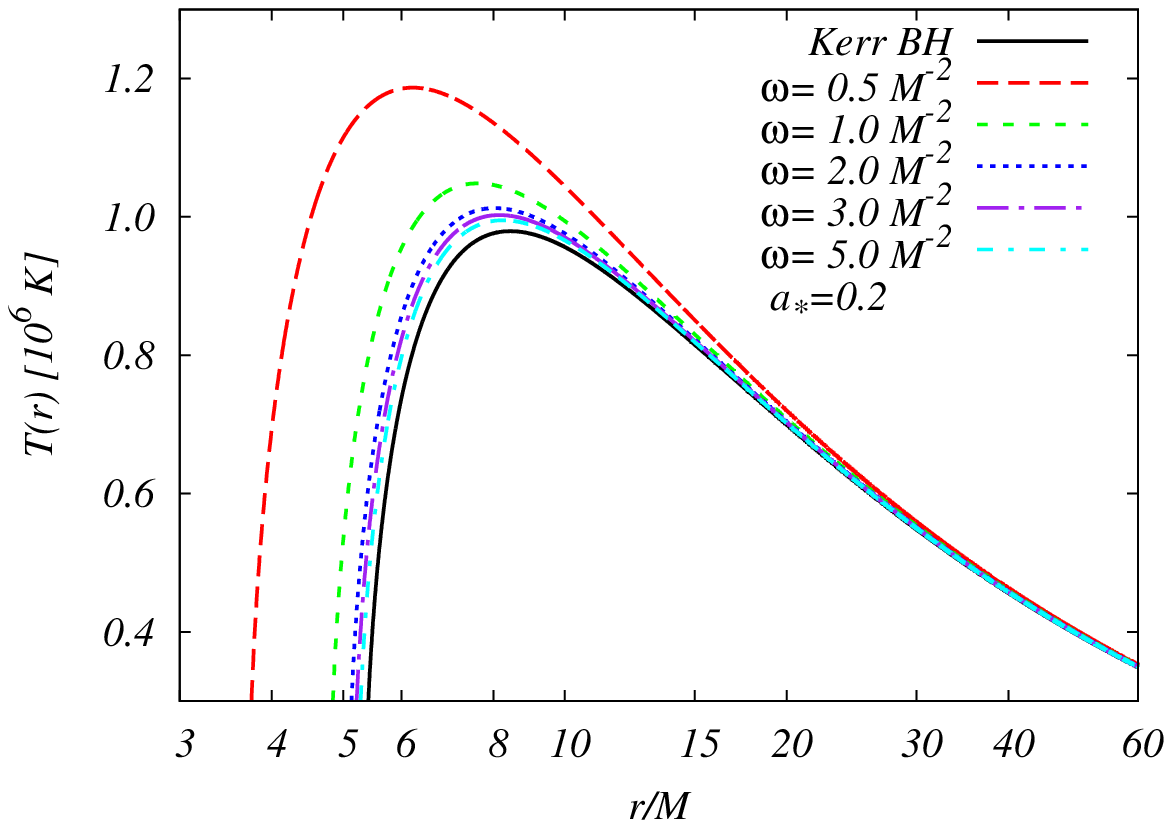}\\
\includegraphics[width=.48\textwidth]{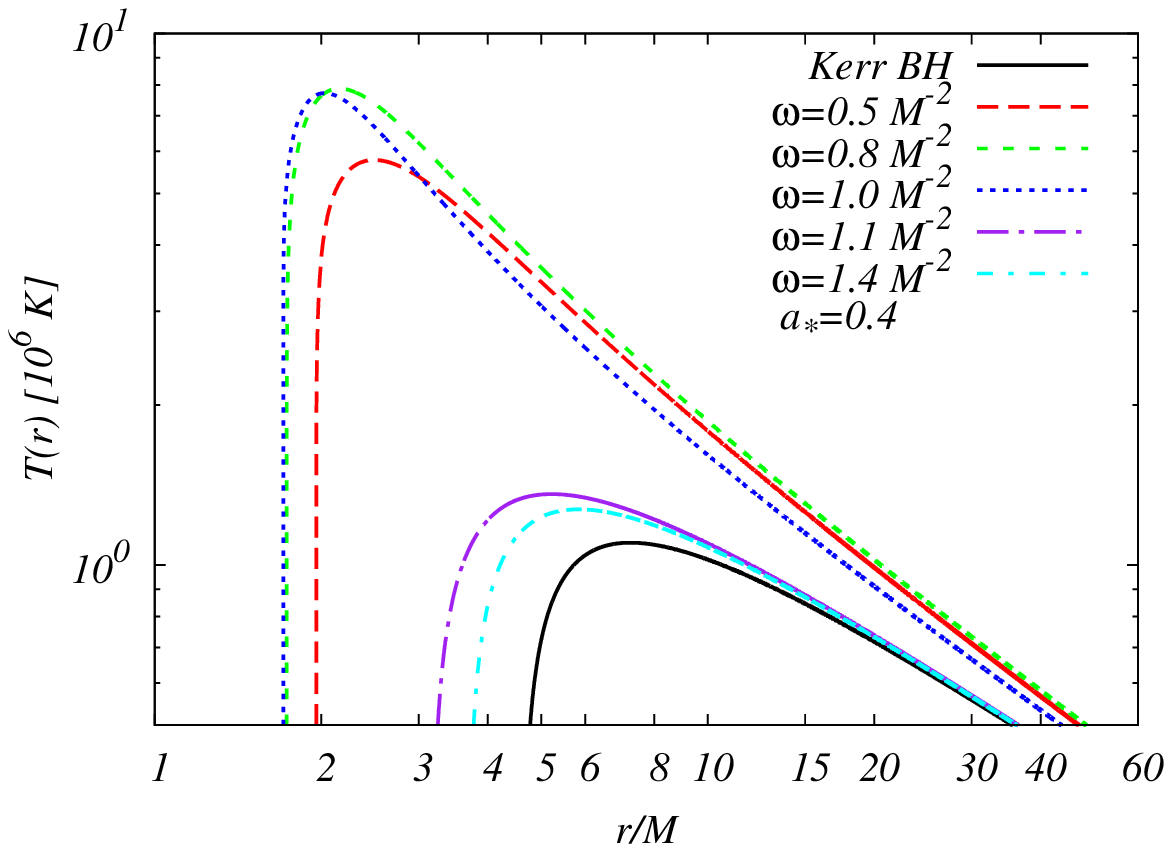}
\includegraphics[width=.48\textwidth]{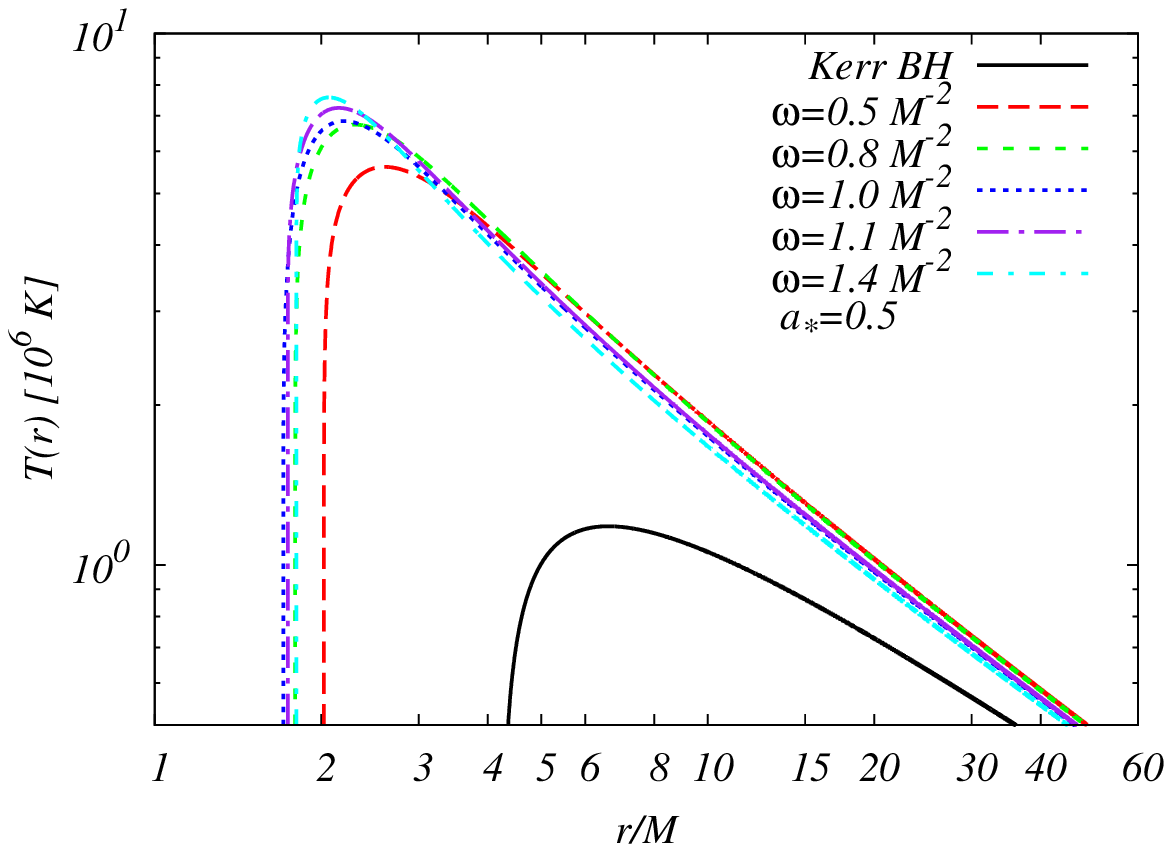}
\caption{The temperature profile of the disk around a static and a slowly rotating KS black hole with the total mass of $1M_{\odot}$ for different values of the parameters $\omega$ and $a_*$. The mass accretion rate is set to $10^{-12}M_{\odot}$/yr.}
\label{fig2}
\end{figure*}

This enhancement in the sensitivity of the variation of $\omega$ can be studied in the radial profiles of the disk temperature as well. In Fig.~\ref{fig2} we present the distribution of the disk temperature for the same values of the parameters $a_*$ and $\omega$ used in Fig.~\ref{fig1}. As compared to the static case, in the case of the configuration with $a_*=0.5$, we see a considerable increase in the disk temperature as a function of $\omega$.

\begin{figure*}[t]
\centering
\includegraphics[width=.48\textwidth]{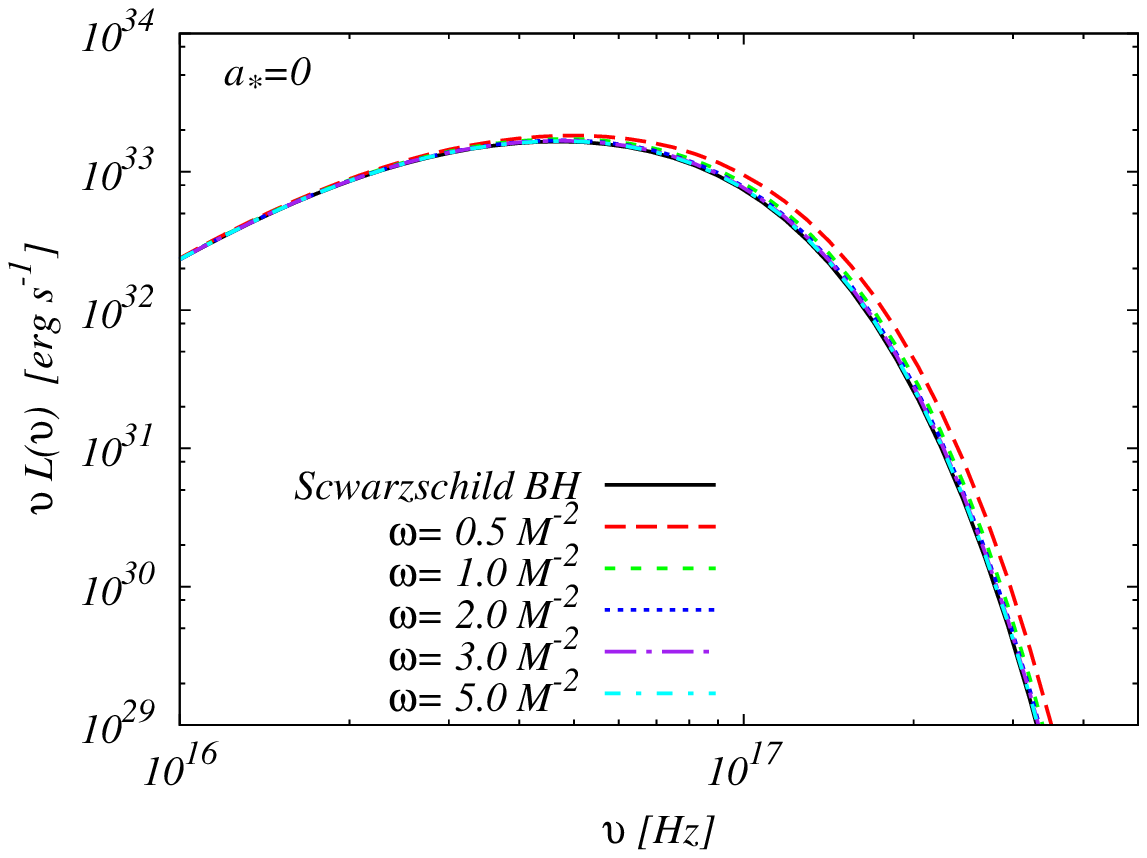}
\includegraphics[width=.48\textwidth]{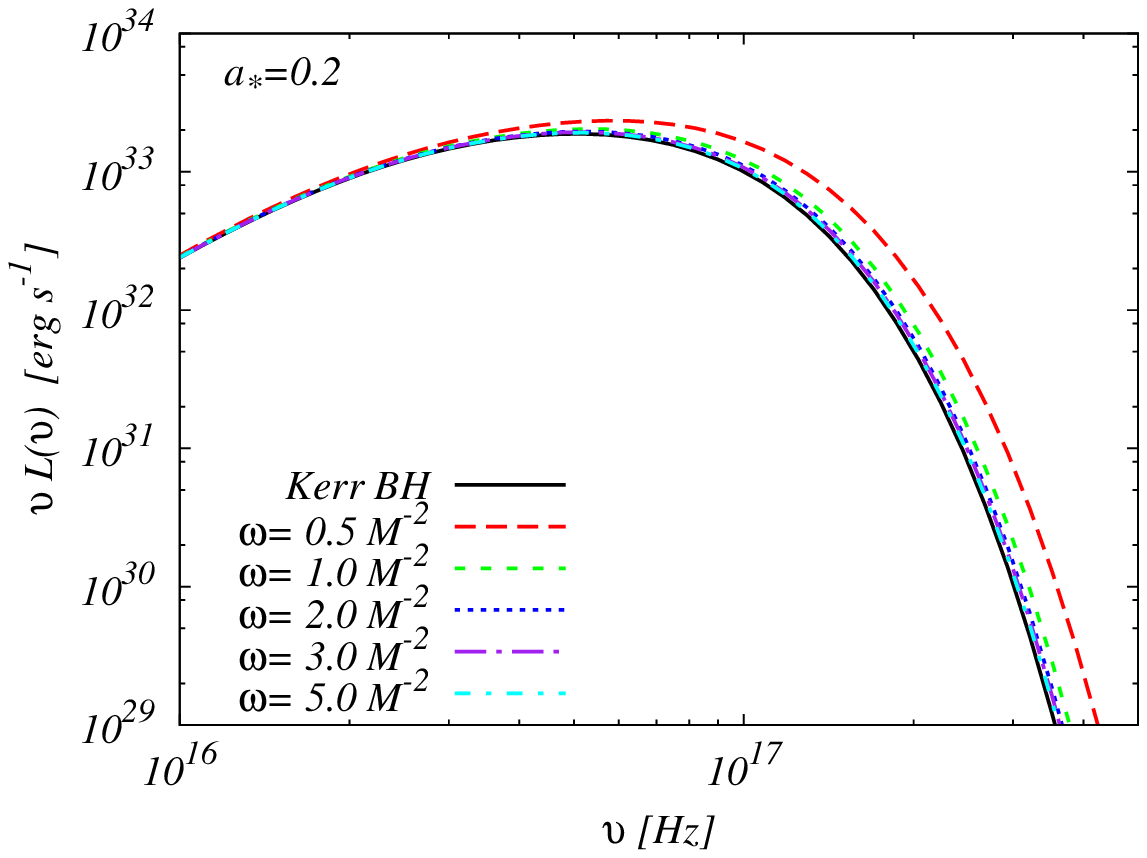}\\
\includegraphics[width=.48\textwidth]{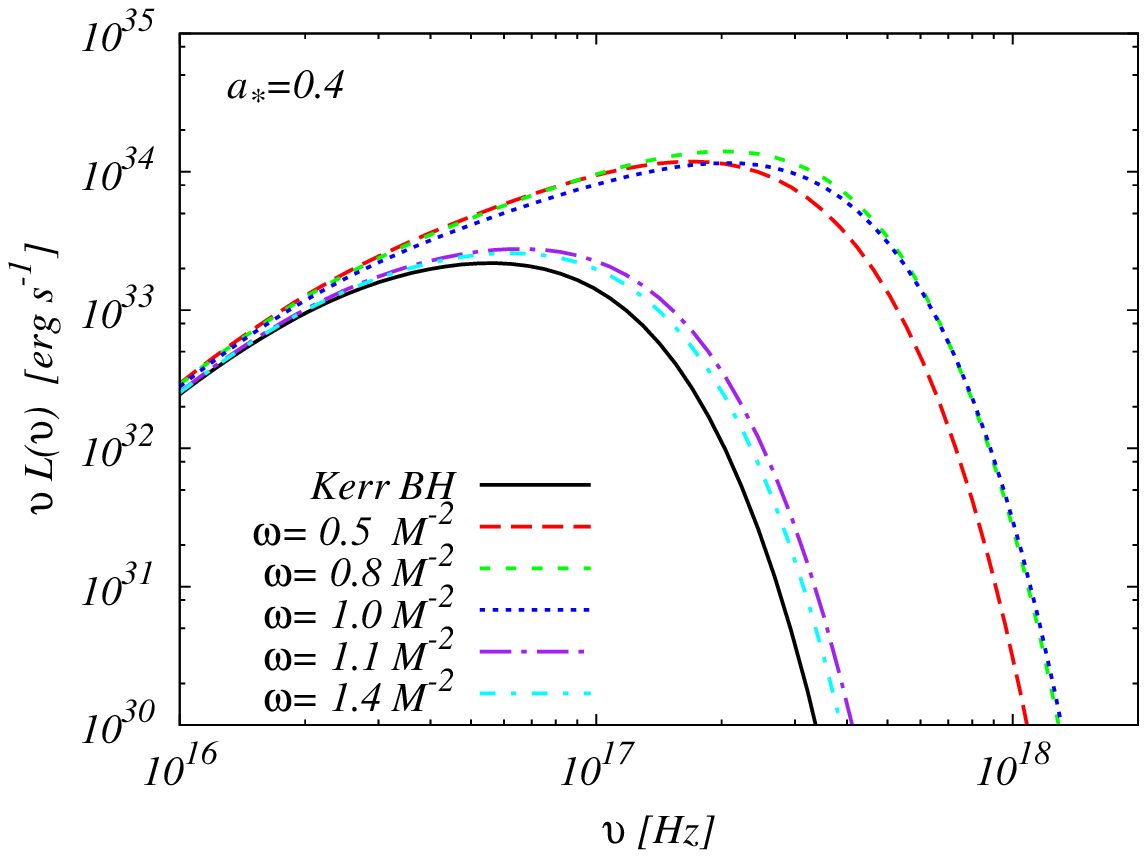}
\includegraphics[width=.48\textwidth]{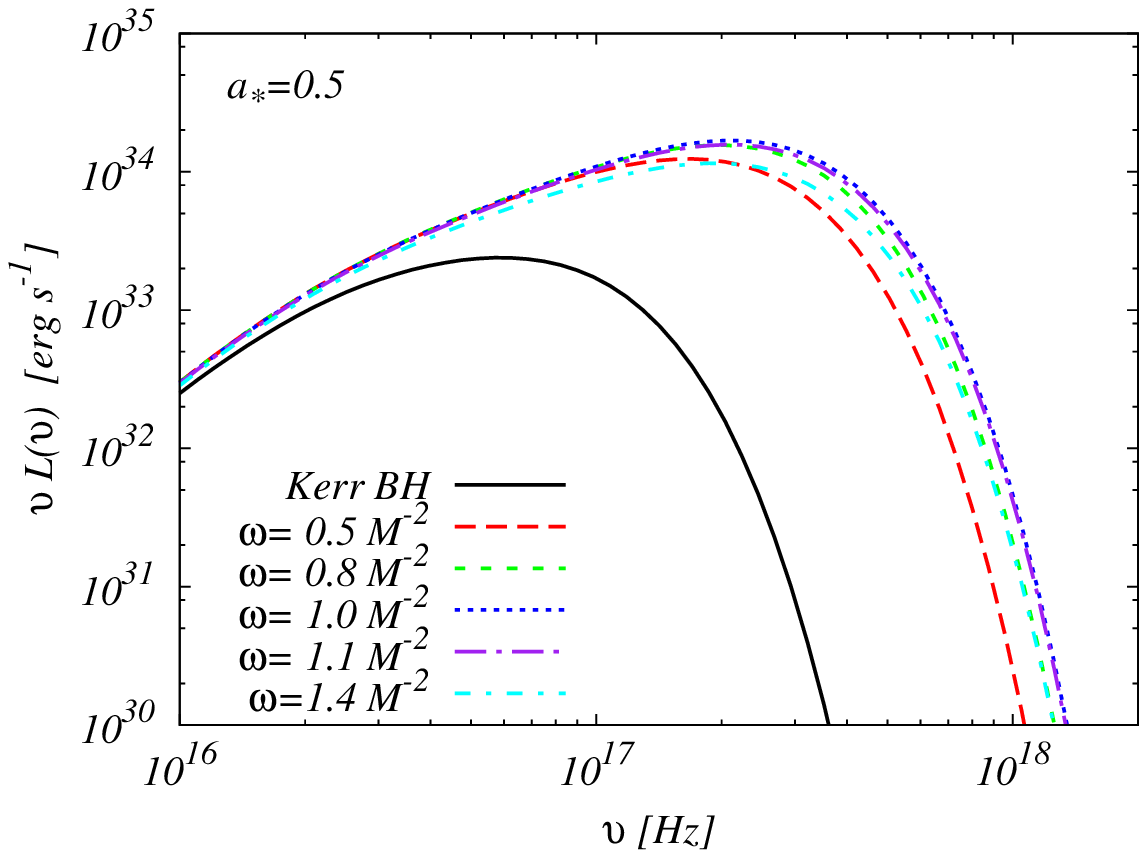}
\caption{The disk spectra for a static and a slowly rotating KS black hole with the total mass of $1M_{\odot}$ for different values of the parameters $\omega$ and $a_*$. The mass accretion rate is set to $10^{-12}M_{\odot}$/yr.}
\label{fig3}
\end{figure*}

For the same set of the parameters $a_*$ and $\omega$ we also plot the disk spectra in Fig.~\ref{fig3}, which were calculated by applying Eq.~(\ref{L}). The trends found in the behavior of the radial distribution of the radiant energy flux and disk temperature can also be seen here. For static spacetimes the disk spectra do not exhibit a strong sensitivity with the variation of $\omega$, and the spectral amplitudes and the cut-off frequency of the spectra have only a moderate increase with decreasing $\omega$. However, for the slowly rotating cases up to $a_*=0.5$, the cut-off frequency can be enhanced by decreasing the value of $\omega$, and the maximal amplitudes have a moderate increase as compared with the general relativistic black holes.

In the cases where for a given $\omega$, $a_*$ exceeds the value of the critical spin,  the differences are considerable, indicating that the accretion disks rotating around a slowly rotating KS black hole have bluer spectra than the disks rotating around a Kerr black hole with identical  mass and spin parameter. Both the increase in the amplitude and the shift in the cut-off frequency is proportional to the radiating surface of the disk. Therefore, they are increased with decreasing $\omega$, provided that the spin is smaller than the critical value corresponding to the given $\omega $. This trend can be seen in the third plot in Fig.~\ref{fig3}, for $a_*=0.4$, when $\omega>M^{-2}$ holds. If the spin exceeds the critical value for a given $\omega$, this trend is
 inverted, as already found in the case of the flux profiles, and the configuration with the minimal value of $\omega$ produces the smallest amplitudes and cut-off frequency of the spectrum. This behavior is shown by the curves obtained for $\omega<M^{-2}$  in the third plot in Fig.~\ref{fig3}, and for each curve obtained for the slowly rotating KS black hole in the last plot in Fig.~\ref{fig3}.

The flux and the emission spectrum of the accretion disks around
compact objects satisfy some  scaling relations, with
respect to the simple scaling transformation of the radial
coordinate, given by $r\rightarrow \widetilde{r}=r/ M$, where $M$
is the mass of the compact sphere. Generally, the metric tensor
coefficients are invariant with respect to this transformation,
while the specific energy, the angular momentum and the angular
velocity transform as
\begin{equation}
 \widetilde{E}\rightarrow \widetilde{E},\qquad
\widetilde{L}\rightarrow M\widetilde{L },
\end{equation}
 and
\begin{equation}
 \Omega
\rightarrow \widetilde{\Omega}/M,
\end{equation}
respectively. The flux scales
as $F(r) \rightarrow F( \widetilde{r})/M^{4}$, giving the simple
transformation law of the temperature as $ T(r)\rightarrow T\left(
\widetilde{r}\right) /M$. By also rescaling the frequency of the
emitted radiation as  $\nu \rightarrow \widetilde{\nu}=\nu /M$,
the luminosity of the disk is given by $L\left( \nu \right)
\rightarrow L\left( \widetilde{\nu}\right) /M$.

Following \cite{PaTh74} we can introduce the scale invariant dimensionless coordinate $x=\sqrt{r/M}$. Then the function $h(r)=\sqrt{1+4M/(\omega r^3)}$ in Eqs.~(\ref{rotE2})-(\ref{rotOmega2}) can be written as $h(x)=\sqrt{1+4/(\tilde{\omega} x^6)}$ where $\tilde{\omega}=\omega M^2$ is a dimensionless parameter. As a consequence, only the specific angular momentum and the rotational frequency have an explicit dependence on $M$ in the form $\widetilde{L}\propto M$ and $\Omega\propto M^{-1}$, whereas $\tilde{E}$ does depend only on $a_*$. For any rescaling $M_2=\alpha M_1$ of the mass of the KS black hole we obtain $x_2 = \alpha^{-1/2} x_1$ and the relations $\widetilde{E}(M_2)(x_2)=\widetilde{E}(M_1)(x_1)$,
$\widetilde{L}(M_2)(x_2)=\alpha\widetilde{L}(M_1)(x_1)$ and
$\Omega(M_2)(x_2)=\alpha^{-1}\Omega(M_1)(x_1)$.
For the flux integral (\ref{F}) and for any rescaling $\dot{M}^{(2)}_0=\beta \dot{M}^{(1)}_0$ of the accretion rate these relations give
\begin{equation}\label{scal1}
F(M_2,\dot{M}^{(2)}_0)(x_2)= (\beta/\alpha^2) F(M_1,\dot{M}^{(1)}_0)(x_1).
\end{equation}
Then the temperature scales as
\begin{equation}\label{scal2}
T(M_2,\dot{M}^{(2)}_0)(x_2)= (\beta/\alpha^2)^{1/4} T(M_1,\dot{M}^{(1)}_0)(x_1).
\end{equation}
For the maximum of the luminosity $L$ we have $\nu_{max}\propto T$ which gives
$L(\nu_{max})\propto \nu^{3}_{max}$. As the frequency scales with the temperature we obtain that
\begin{equation}\label{scal3}
L(M_2,\dot{M}^{(2)}_0)(\nu_2)= (\beta/\alpha^2)^{3/4} L(M_1,\dot{M}^{(1)}_0)(\nu_1),
\end{equation} 
with 
\begin{equation}\label{scal4}
\nu_2(M_2,\dot{M}^{(2)}_0)= (\beta/\alpha^2)^{1/4} \nu_1(M_1,\dot{M}^{(1)}_0).
\end{equation}
With the help of these scaling relations we can always obtain the values of the flux and of the luminosity for an arbitrary value of the mass, once these values are known for a given mass.

In Table~\ref{tab1} we present the conversion efficiency $\epsilon$ of the accreted mass into radiation for both the Kehagias-Sfestos and general relativistic black holes. As seen in Eq.~(\ref{epsilon}), the value of $\epsilon$ highly depends on the location of the inner edge of the disk. Under the critical value of the spin,  $r_{ms}$ shifts to lower radii for lower values of $\omega$, and the efficiency obtained for the rotating Kehagias-Sfestos case is increasing as compared with the efficiency of  the general relativistic black holes. By decreasing the value of $\omega$, the efficiency increases, and the degree of this increase is enhanced for the black holes rotating at higher spin. If the latter is below the critical value of the spin,  then this enhancement is inversely proportional to $\omega$.

Above the critical value of the spin the inner edge of the disk is located at the event horizon, where $g_{tt}=0$. In Fig.~\ref{fig00} we have also shown that $\Omega$ falls to zero at $r_{+}$. Then the specific energy given by Eq.~ (\ref{rotE}) vanishes at the horizon. Since $E_{ms}$ is zero at the inner edge of the disk, the particles deposited on the black hole carry only their rest mass energy into the black hole, which is negligible as compared to the total thermal energy radiated by the disk. As a result, $\epsilon$ would essentially be one, i.e, the efficiency is 100\% for spin values above the critical spin. However, we have also seen that the zone close to the inner edge of the disk cannot be cooled by thermal radiation only. Therefore in this case the application of the standard thin disk model is problematic. At its inner edge the disk may become geometrically thick, and other forms of the stress-energy, besides the specific energy and the specific angular momentum can play an important role in the mass-energy deposition. Then the usage of Eq.~(\ref{epsilon}) becomes problematic. We can still consider that, although the area of the forbidden radiative cooling, located between $r_{+}$ and $r_{max}$ (the radius where $\Omega$ is maximal), is big enough to cause measurable effects in the energy-angular momentum transport between the black hole and the disk.  However, this transport may still not determine drastic changes in the standard thin disk model. Hence, we expect that in this region the efficiency could still be very high, though not 100\%.

Therefore, we conclude that the Kehagias-Sfetsos black holes convert accreted mass to radiation in a more efficient way than general relativistic black holes do. This enhancement in efficiency occurs even if the disk located around the rotating KS black holes can emit thermal photons only over some part of its entire surface area.

\begin{table}[tbp]
\begin{center}
\begin{tabular}{|c|c|c|c|c|}
\hline
$\omega/M^{2}$ & $a_*=0$ & $0.2$ & $0.4$ & $0.5$ \\
\hline
1/2 & 6.30 (5.28) & 8.07 (3.72) & $\sim\!\!100$ (1.71)  & $\sim\!\!100$ (1.00) \\
1 & 5.97 (5.66) & 7.00 (4.78) & $\sim\!\!100$ (1.71) &  $\sim\!\!100$ (1.00) \\
2 & 5.83 (5.84) & 6.73 (5.06) & 8.45 (3.97) & $\sim\!\!100$ (1.00) \\
3 & 5.79 (5.89) & 6.65 (5.14) & 8.20 (4.15) & 9.92 (3.35)  \\
5 & 5.76 (5.94) & 6.59 (5.21)& 8.03 (4.26) & 9.46 (3.59) \\
10 & 5.74 (6.00) & 6.54 (5.29) & 7.92 (4.37) &  9.20 (3.73) \\
100 & 5.72 (6.00) & 6.51 (5.29) & 7.83 (4.45) & 9.01 (3.84) \\
$\infty$ & 5.72 (6.00) & 6.46 (5.33) & 7.51 (4.62) & 8.21 (4.24) \\
\hline
\end{tabular}
\end{center}
\caption{The efficiency (measured in percents), and the values of the radius divided by $M$ at the inner edge  (values in parenthesis) of the accretion disk for slowly-rotating black holes in general relativity, and in the
HL modified theory of gravity with the rotating KS solution. The case of $\omega\rightarrow\infty$ corresponds to the general relativistic case.} \label{tab1}
\end{table}

\section{Discussions and final remarks}
\label{sec:concl}

In the present paper, we have studied thin accretion disk models
for the slowly rotating Kehagias and Sfetsos black hole solution in Ho\v{r}ava gravity, and we have carried out an analysis of the properties of the radiation emerging from the surface of the disk.  
By comparing the accretion disk properties in the  slowly rotating geometry in the  Ho\v{r}ava-Lifshitz gravity with the properties of disks around a slowly rotating Kerr black hole, we have shown that the intensity of the flux emerging from the disk surface is greater for the
slowly rotating Kehagias and Sfetsos solution than for the general relativistic rotating Kerr black hole with the same geometrical mass $r_0$ and accretion rate $\dot{M}_0$. If this difference is in the range of 20\%-80\% for the case of slowly rotating black holes, in the case of KS black holes with $a_*$ of the order of $a_*=0.4-0.5$, the increase in the flux could be several orders of magnitude higher than for a Kerr black hole. Thus extreme high energy emissions from accretion disks around black hole candidates may provide the distinctive signature for a KS black hole. We have also
presented the conversion efficiency $\epsilon$ of the accreting
mass into radiation, and we have showed that the slowly rotating Kehagias and Sfetsos black holes are much more efficient in converting the accreting mass into radiation than their Kerr black holes counterparts. 

From an observational point of view, the determination of the mass and of the spin of the central compact general relativistic object is a very difficult task, and  in many cases, their values are not known. Usually what is observed is the flux and the spectrum of the radiation coming from the accretion disk. Therefore, an essential problem  that has to be considered is what properties of the radiation spectrum and flux, respectively,  could indicate that the observed black hole is a Kehagias and Sfetsos black hole of the Ho\v{r}ava-Lifshitz gravity, rather than a standard Kerr black hole, with a different mass and spin. With the use of Eqs.~(\ref{scal1})-(\ref{scal4}) we can always describe the properties of the accretion disk in terms of an "effective" Kerr black hole, with a given "effective" mass and spin, respectively. Since for a Kehagias and Sfetsos black hole the emitted energy flux could be several orders of magnitude higher than for a Kerr black hole, for a given maximal value of $a_{\ast }$, the observed high flux from the Kehagias and Sfetsos black hole would require an unrealistically high mass for the central general relativistic compact object. Moreover, for the Kehagias and Sfetsos black hole the maximum of the flux is displaced toward smaller values of $\nu /M$, as compared to the Kerr case. Hence a very high Kerr mass of the central object, and the position of the maximum of the spectrum could represent an important observational evidence for the existence of the Kehagias and Sfetsos black holes.

Therefore, the study of the accretion processes by compact objects is a powerful indicator of their physical nature. Since the energy flux, the temperature distribution of the disk, the spectrum  of the emitted black body radiation, as well as the conversion efficiency show, in the case of the Ho\v{r}ava-Lifshitz theory vacuum solutions, significant differences as compared to the general relativistic case, the determination of these
observational quantities could discriminate, at least in principle, between standard general relativity and Ho\v{r}ava-Lifshitz theory, and constrain the parameters of the model.

\section*{Acknowledgments}

The work of TH is supported by an RGC grant of the government of the Hong Kong SAR. FSNL acknowledges financial support of the Funda\c{c}\~{a}o para a Ci\^{e}ncia e Tecnologia through the grants PTDC/FIS/102742/2008 and CERN/FP/109381/2009.

\appendix
\section{Physical properties of thin accretion disks}\label{App1}

For a thin accretion disk the vertical size (defined in
cylindrical coordinates along the $z$-axis) is negligible, as
compared to its horizontal extension (defined along the radial
direction $r$), i.e., the disk height $H$, equal to the maximum
half thickness of the disk, is always much smaller than the
characteristic radius $R$ of the disk, $H \ll R$.  The thin disk
is assumed to be in hydrodynamical equilibrium, and the pressure
gradient, as well as the vertical entropy gradient, are negligible
in the disk. The efficient cooling via the radiation over the disk
surface prevents the disk from cumulating the heat generated by
stresses and dynamical friction. In turn, this equilibrium causes
the disk to stabilize its thin vertical size. The thin disk has an
inner edge at the marginally stable orbit of the compact object
potential, and the accreting matter has a Keplerian motion in
higher orbits.

In steady-state accretion disk models, the mass accretion rate
$\dot{M}_{0}$ is assumed to be a constant that does not change
with time. The physical quantities describing the orbiting matter
are averaged over a characteristic time scale, e.g. $\Delta t$,
for a total period of the orbits, over the azimuthal angle $\Delta
\phi =2\pi $,  and over the height $H$
\cite{ShSu73,NoTh73,PaTh74}.

The particles moving in Keplerian orbits around the compact object
with a rotational velocity $\Omega =d\phi /dt$ have a specific
energy $\widetilde{E} $ and a specific angular momentum
$\widetilde{L\text{,}}$ which in the steady-state thin disk model
depend only on the radii of the orbits. These particles, orbiting
with the four-velocity $u^{\mu }$, form a disk of an averaged
surface density $\Sigma $, the vertically integrated average of
the rest mass density $\rho _{0}$ of the plasma. The accreting
matter in the disk is modeled by an anisotropic fluid source,
where the density $\rho _{0}$, the energy flow vector $q^{\mu }$
and the stress tensor $t^{\mu \nu }$ are measured in the averaged
rest-frame (the specific heat was neglected). Then, the disk
structure can be characterized by the surface density of the disk
\cite{NoTh73,PaTh74}
\begin{equation}
\Sigma(r) = \int^H_{-H}\langle\rho_0\rangle dz,
\end{equation}
with averaged rest mass density $\langle\rho_0\rangle$ over
$\Delta t$ and $ 2\pi$ and the torque
\begin{equation}
W_{\phi}{}^{r} =\int^H_{-H}\langle t_{\phi}{}^{r}\rangle dz,
\end{equation}
with the averaged component $\langle t^r_{\phi} \rangle$ over
$\Delta t$ and $2\pi$. The time and orbital average of the energy
flux vector gives the radiation flux $F(r)$ over the
disk surface as $F(r)=\langle q^z \rangle$.

The stress-energy tensor is decomposed according to
\begin{equation}
T^{\mu \nu }=\rho_{0}u^{\mu }u^{\nu }+2u^{(\mu }q^{\nu )}+t^{\mu
\nu },
\end{equation}
where $u_{\mu }q^{\mu }=0$, $u_{\mu }t^{\mu \nu }=0$. The
four-vectors of the energy and angular momentum flux are defined
by $-E^{\mu }\equiv T_{\nu }^{\mu }{}(\partial /\partial t)^{\nu
}$ and $J^{\mu }\equiv T_{\nu }^{\mu }{}(\partial /\partial \phi
)^{\nu }$, respectively. The structure equations of the thin disk
can be derived by integrating the conservation laws of the rest
mass, of the energy, and of the angular momentum of the plasma,
respectively \cite{NoTh73,PaTh74}. From the equation of
the rest mass conservation, $\nabla _{\mu }(\rho _{0}u^{\mu })=0$,
 it follows that the time averaged rate of the accretion of the
rest mass is independent of the disk radius,
\begin{equation}
\dot{M_{0}}\equiv -2\pi r\Sigma u^{r}={\rm constant}.
\label{conslawofM}
\end{equation}

Applying the conservation law $\nabla _{\mu }E^{\mu }=0$ of the energy has
the integral form
\begin{equation}
\lbrack \dot{M}_{0}\widetilde{E}-2\pi r\Omega W_{\phi }{}^{r}]_{,r}
=4\pi r  F%
\widetilde{E},  \label{conslawofE}
\end{equation}
which states that the energy transported by the rest mass flow,
$\dot{M}_{0} \widetilde{E}$, and the energy transported by the
dynamical stresses in the disk, $2\pi r\Omega W_{\phi }{}^{r}$, is
in balance with the energy radiated away from the surface of the
disk, $4\pi rF\widetilde{E}$. The law of the angular
momentum conservation, $\nabla _{\mu }J^{\mu }=0$, also states the
balance of these three forms of the angular momentum transport,
\begin{equation}
\lbrack \dot{M}_{0}\widetilde{L}-2\pi rW_{\phi }{}^{r}]_{,r}=4\pi
r F \widetilde{L}.  \label{conslawofL}
\end{equation}

By eliminating $W_{\phi }{}^{r}$ from Eqs.~(\ref{conslawofE}) and
(\ref {conslawofL}), and applying the universal energy-angular
momentum relation $ dE=\Omega dJ$ for circular geodesic orbits in
the form $\widetilde{E}_{,r}=\Omega \widetilde{L}_{,r}$, the flux
$F$ of the radiant energy over the disk can be
expressed in terms of the specific energy, angular momentum and of
the angular velocity of the compact sphere \cite{NoTh73,PaTh74},
\begin{equation}
F(r)=-\frac{\dot{M}_{0}}{4\pi \sqrt{-g}}\frac{\Omega
_{,r}}{(\widetilde{E}-\Omega
\widetilde{L})^{2}}\int_{r_{ms}}^{r}(\widetilde{E}-\Omega
\widetilde{L}) \widetilde{L}_{,r}dr.  \label{F}
\end{equation}

Another important characteristic of the mass accretion process is
the efficiency with which the central object converts rest mass
into outgoing radiation. This quantity is defined as the ratio of
the rate of the radiation of the energy of photons escaping from
the disk surface to infinity, and the rate at which mass-energy is
transported to the central compact general relativistic object,
both measured at infinity \cite{NoTh73,PaTh74}. Since the former
is given by the rate of the difference between the total mass growth of the central object and the accreted rest mass, the efficiency can be written as
$\epsilon = (\Delta M - \Delta M_0)/\Delta M_0$.
The accretion of the rest mass $\Delta M_0$ producing a growth in the total mass-energy of the central object proportional to the specific energy of the
gas particles leaving the marginally stable orbit, i.e
 $\Delta M = \widetilde{E}_{ms}\Delta M_0$.
Therefore the efficiency is given in terms of the specific energy measured at the marginally stable orbit
$r_{ms}$,
\begin{equation}
\epsilon =1-\widetilde{E}_{ms},\label{epsilon}
\end{equation}
provided all the emitted photons can escape to infinity.

For Schwarzschild black holes the efficiency $\epsilon $ is about
$6\%$, whether the photon capture by the black hole is considered,
or not. Ignoring the capture of radiation by the hole, $\epsilon $
is found to be $42\%$ for rapidly rotating black holes, whereas
the efficiency is $40\%$ with photon capture in the Kerr potential
\cite{Th74}.

The accreting matter in the steady-state thin disk model is
supposed to be in thermodynamical equilibrium. Therefore the
radiation emitted by the disk surface can be considered as a
perfect black body radiation, where the energy flux is given by
$F (r)=\sigma T^{4}(r)$ ($\sigma $ is the
Stefan-Boltzmann constant), and the observed luminosity $L\left(
\nu \right) $ has a redshifted black body spectrum \citep{To02}:
\begin{equation}
L\left( \nu \right) =4\pi d^{2}I\left( \nu \right)
=\frac{8}{\pi }\cos \gamma \int_{r_{i}}^{r_{f}}
\int_0^{2\pi}\frac{\nu^{3}_e r d\phi dr }{\exp \left( h\nu_e/T\right) -1}.\label{L}
\end{equation}

Here $d$ is the distance to the source, $I(\nu )$ is the Planck
distribution function, $\gamma $ is the disk inclination angle,
and $r_{i}$ and $r_{f}$ indicate the position of the inner and
outer edge of the disk, respectively. We take $r_{i}=r_{ms}$ and
$r_{f}\rightarrow \infty $, since we expect the flux over the disk
surface vanishes at $r\rightarrow \infty $ for any kind of general
relativistic compact object geometry. The emitted frequency is
given by $\nu_e=\nu(1+z)$, and the redshift factor can be written
as
\begin{equation}
1+z=\frac{1+\Omega r \sin \phi \sin \gamma }{\sqrt{ -g_{tt}
- 2 \Omega g_{t\phi} - \Omega^2 g_{\phi\phi}}},
\end{equation}
where we have neglected the light bending \citep{Lu79,BMT01}.


\end{document}